\documentclass[12pt]{article}
\def\beq{\begin{equation}}
\def\eeq{\end{equation}}

\begin{document}
\begin{titlepage}
\begin{center}
{\Large \bf William I. Fine Theoretical Physics Institute \\
University of Minnesota \\}  \end{center}
\vspace{0.2in}
\begin{flushright}
FTPI-MINN-04/38 \\
UMN-TH-2324-04 \\
October 2004 \\
\end{flushright}
\vspace{0.3in}
\begin{center}
{\Large \bf  A possible gateway to $\eta_b$: $\chi_{b0}(2P) \to \eta \, \eta_b$
\\}
\vspace{0.2in}
{\bf M.B. Voloshin  \\ }
William I. Fine Theoretical Physics Institute, University of
Minnesota,\\ Minneapolis, MN 55455 \\
and \\
Institute of Theoretical and Experimental Physics, Moscow, 117259
\\[0.2in]
\end{center}

\begin{abstract}

It is argued that the branching ratio for the decay $\chi_{b0}(2P) \to \eta \,
\eta_b$ can reach few permil as a result of the enhancement of the $\eta$
emission by the axial anomaly in QCD. This might make the discussed process
practical for an experimental search for the $\eta_b$. 

\end{abstract}

\end{titlepage}

The ground $1^1S_0$ state of bottomonium, the pseudoscalar $\eta_b$ resonance,
is expected to have mass approximately $40 \pm 10$ MeV below that of the
$\Upsilon$ resonance. An early QCD calculation\cite{mv} has estimated the
hyperfine splitting at about 35 MeV, while the latest perturbative QCD
calculation\cite{kppss} in the next-to-next-to-leading order provides the
central estimated value of the splitting as 39 MeV with a theoretical
uncertainty of about 10 MeV (and an additional uncertainty due to the current
knowledge of the QCD coupling $\alpha_s$). The preliminary lattice
calculations\cite{ukqcd} favor a somewhat larger value of the splitting, up to
approximately 50 MeV. Thus an experimental observation of the $\eta_b$
resonance, besides being of a general interest for the study of the heavy
quarkonium, would provide an important input into the current theoretical
approaches.

It is however known that experimentally the $\eta_b$ is quite elusive: the rate
of the allowed M1 radiative transition $\Upsilon \to \gamma \, \eta_b$ is
greatly suppressed by the small value of the hyperfine splitting, while
analogous transitions from the excited $^3S_1$ bottomonium resonances,
$\Upsilon(2S) \to \gamma \, \eta_b$ and $\Upsilon(3S) \to \gamma \, \eta_b$, are
forbidden in the nonrelativistic limit by the vanishing overlap of the wave
functions. Thus the latter transitions proceed only due to the relativistic
effects, which are quite small in bottomonium and are highly model-dependent.
The purpose of the present letter is to point out that the decay chain
$\Upsilon(3S) \to \gamma \, \chi_{b0}(2P)$ followed by $\chi_{b0}(2P) \to \eta
\, \eta_b$ can provide an alternative realistic approach to an experimental
search for the $\eta_b$. The first transition in this chain is well
known\footnote{The PDG\cite{pdg} value for the branching ratio $B(\Upsilon(3S)
\to \gamma \, \chi_{b0}(2P))$ is $5.4 \pm 0.6$\%, while the latest CLEO
measurement\cite{cleo} yields $6.77 \pm 0.20 \pm 0.65$\%.} and is a reliable
source of the $\chi_{b0}(2P)$ resonance. (The CLEO data sample
contains\cite{cleo} $(225 \pm 7)\times 10^3$ observed events with the decay
$\Upsilon(3S) \to \gamma \, \chi_{b0}(2P)$.) It will be argued here that the
branching ratio for the decay $\chi_{b0}(2P) \to \eta \, \eta_b$, although still
uncertain, may be as large as few permil, due to a considerable enhancement of
the $\eta$ emission in the quarkonium transition by the axial anomaly in QCD.
Depending on the experimental technique, the background conditions for the
discussed here decay chain may be more favorable than in a search for the
$\eta_b$ by the direct radiative transition from $\Upsilon(3S)$.

The amplitude of the discussed here $\eta$ transition in bottomonium carries the
usual suppression corresponding to breaking of the flavor SU(3) as well as a
suppression by the factor $m_b^{-1}$ corresponding to the spin-flip of the heavy
$b$ quark. However these factors are to some extent compensated by that the
considered decay is an $S$ wave process and by the enhancement of the $\eta$
production by soft gluonic field, expressed by the relation\cite{aanom}
\beq
\langle \eta | G^a {\tilde G}^a |0 \rangle = 8 \pi^2 \, \sqrt{2 \over 3}
\, f_\eta \, m_\eta^2~,
\label{eanom}
\eeq
which is a consequence of the anomaly in the SU(3) flavor singlet axial current
in QCD\footnote{The normalization of the gluonic field used here includes the
QCD coupling $g$, which normalization corresponds to writing the gluon
Lagrangian as $L_g=-G^2/(4 g^2)$.}. The constant $f_\eta$ is the annihilation
constant for the $\eta$ meson, analogous to the pion constant $f_\pi \approx
130$ MeV. One can also notice that eq.(\ref{eanom}) contains the flavor SU(3)
breaking factor $m_\eta^2 \propto m_s$.

The relation of the amplitudes of hadronic transitions in heavy quarkonium to
the matrix elements of the type shown in eq.(\ref{eanom}) arises within the
description\cite{gottfried} of the hadronic transitions in terms of the
multipole expansion in QCD\cite{gottfried,mv2}. Within this approach the heavy
quarkonium is considered as a compact object whose interaction with the soft
gluonic fields can be expanded in multipoles, while the production of the light
hadrons in the transition is described\cite{vz,mv3} by matrix elements analogous
to that in eq.(\ref{eanom}). However there is a serious uncertainty within this
approach associated with evaluation of the heavy quarkonium transition
amplitudes resulting from the interactions with gluonic field. This is
especially true for transitions from the states whose radial wave function has
nodes, such as the discussed here transition $2P \to 1S$, where considerable
cancellations in the overlap integral do take place. In lieu of a better
model-independent approach, another known $2P \to 1S$ transition, $\chi_{b0}(2P)
\to \gamma \, \Upsilon$, is used here for an estimate of the scale of the
corresponding matrix element.

The decay $\chi_{b0}(2P) \to \eta \, \eta_b$ is induced by the interference of
the E1 and M1 terms in the multipole expansion in QCD. The Hamiltonian
describing these two terms can be written as
\beq
H_{E1}+H_{M1}=-{1 \over 2} \xi^a \, {\vec r} \cdot {\vec E}^a (0) - {1 \over 2
\, m_b}\, \xi^a \, ({\vec s}_1-{\vec s}_2) \cdot {\vec B}^a(0)~,
\label{ham}
\eeq
where $\xi^a=t_1^a-t_2^a$ is the difference of the color generators
acting on the quark and antiquark, ${\vec s}_1-{\vec s}_2$ is the difference of
the corresponding spin operators, and ${\vec r}$ is the vector
for relative position of the quark and the antiquark. Finally, ${\vec E}^a$ and
${\vec B}^a$ are the chromoelectric and the chromomagnetic components of the
gluon field strength.

In a nonrelativistic quarkonium the spin, orbital, and the radial degrees of
freedom factorize. Thus using the Hamiltonian in eq.(\ref{ham}) in the second
order and retaining only the relevant interference term, one can readily find
the  amplitude of the discussed transition in the form
\beq
A(\chi_{b0}(2P) \to \eta \, \eta_b)={1 \over 92 \, m_b} \, \langle \eta | {\vec
E}^a \cdot {\vec B}^a | 0 \rangle \, J={\pi^2 \over 48} \, \sqrt{2 \over 3} \,
{f_\eta \, m_\eta^2 \over m_b} \, J~,
\label{ampe}
\eeq
where the axial anomaly relation (\ref{eanom}) is taken into account, and $J$ is
the radial part of the quarkonium transition amplitude, depending on the overlap
of the radial wave functions $R_{2P}$ and $R_{1S}$ as
\beq
J=\langle R_{1S} |\, r \, \xi^c \, {\cal G}_P\, \xi^c \,| R_{2P} \rangle +
\langle \, R_{1S} |\, \xi^c \, {\cal G}_S\, \xi^c \, r \, | R_{2P} \rangle~.
\label{ampj}
\eeq
Here ${\cal G}$ stands for the Green function of the heavy quarkonium in the
color octet state. Clearly the first term contains its $P$ wave part, while the
second term contains the $S$ wave part of ${\cal G}$. At present it is still not
clear how to calculate this Green function, which is the main source of
uncertainty in estimating the absolute rates of the hadronic transitions. It is
believed to be mainly contributed by the states above the open flavor
threshold\cite{mv2}. In this situation for an estimate one can replace this
Green function by a local operator ${\cal G} \to 1/\Delta$ with $\Delta$ being
the ``effective energy gap" to the states contributing to ${\cal G}$. Such
approximation is in a reasonable agreement with the data on the transitions
$\psi(2S) \to \pi \pi \, J/\psi$ and $\Upsilon(2S) \to \pi \pi \, \Upsilon$,
with $\Delta \sim 1$ GeV. Adopting this approximation and taking into account
the color factor $\xi^a \xi^a = 16/3$, the expression for the amplitude $J$ in
eq.(\ref{ampj}) can be greatly simplified:
\beq
J \approx {32 \over 3 } \, {I \over  \Delta}
\label{ampj1}
\eeq
with $I$ given by
\beq
I=\langle R_{1S} |\, r  \,| R_{2P} \rangle~.
\label{ampi}
\eeq
One can notice that the overlap integral $I$ describes the matrix element for
the electric dipole transition\footnote{The form factor effects are neglected
here for both the hadronic and the radiative transitions. The corrections due to
those effects are smaller than the main uncertainty in the discussed calculation
related to the Green function ${\cal G}$.} $\chi_{b0}(2P) \to \gamma \,
\Upsilon$, and that the rate of the latter transition is given by the well known
formula
\beq
\Gamma(\chi_{b0}(2P) \to \gamma \, \Upsilon) = {4 \over 9} \, \alpha \, Q_b^2 \,
\omega_\gamma^3 \, |I|^2~,
\label{rgam}
\eeq
where $Q_b=-1/3$ is the electric charge of the $b$ quark and $\omega_\gamma
\approx 0.74$ GeV is the energy of the emitted photon.

Using the equations (\ref{ampe}) and (\ref{ampj1}) one can thus relate the rates
of the two transitions as
\beq
{\Gamma(\chi_{b0}(2P) \to \eta \, \eta_b) \over \Gamma(\chi_{b0}(2P) \to \gamma
\, \Upsilon)} \approx {\pi^3 \over 3 \, \alpha} \, {p_\eta \, f_\eta^2 \,
m_\eta^4 \over \omega_\gamma^3 \, m_b^2 \, \Delta^2} \approx 0.2 \left ( {f_\eta
\over 0.16 ~{\rm GeV}} \right)^2 \, \left( {1~{\rm GeV} \over \Delta} \right )^2
\label{ratg}
\eeq
with $p_\eta$ being the momentum of the emitted $\eta$ meson.
According to PDG\cite{pdg} the branching ratio for the radiative decay is
$B(\chi_{b0}(2P) \to \gamma \, \Upsilon) = 0.9 \pm 0.6 \%$. If the central value
of the current data can serve as a reasonable guideline, the branching ratio of
the discussed transition $\chi_{b0}(2P) \to \eta \, \eta_b$ can thus be
estimated as likely exceeding $10^{-3}$, at which level the transition is
hopefully within the reach of the experiment.

I thank Dan Cronin-Hennessy for most useful discussions of the possibilities for
an experimental search of the discussed here decay. This work is supported in
part by the DOE grant DE-FG02-94ER40823.

\end{document}